# High-Efficiency Broadband Mid-Infrared Absorption in Asymmetrically Matched Metallic Meanders: Development of $Ti_{40}V_{60}$ Alloy based LEKIDs for Mid-Far IR

Shilpam Sharma[1], Shekhar Chandra Pandey[1,2], Anudeep Singh[1], Shankar Lal[1,2], R. S. Saini[1] and M. K. Chattopadhyay[1,2]

[1]*Free Electron Laser Utilization Section, Raja Ramanna Centre for Advanced Technology, Indore, Madhya Pradesh - 452 013, India*

[2]*Homi Bhabha National Institute, Training School Complex, Anushakti Nagar, Mumbai 400 094, India*

**Abstract.** Fast, highly sensitive, and broadband detectors operating in the mid-to-far-infrared (MIR-FIR) spectral region are essential for applications ranging from astrophysics to time-resolved spectroscopy using laboratory-based sources and beamlines at Infrared Free-Electron Laser (IR-FEL) facilities. Conventional Lumped Element Kinetic Inductance Detectors (LEKID) achieve high optical efficiency using resonant quarter wavelength (λ/4) backshort cavities. However, this cavity-based approach is inherently narrowband and becomes optomechanically challenging at the mid to far infrared (MIR-FIR) wavelengths. Here, we present a completely backshort free LEKID absorber architecture that achieves broadband absorption exceeding 50%. The meander absorbers were fabricated out of superconducting Ti-V alloy and characterized using the IR radiation from the IR-FEL at RRCAT, India. By illuminating through the silicon substrate and matching the sheet resistance of the meander to the wave impedance of the silicon substrate, the front-side reflection is strongly suppressed. Simultaneously, the sub-wavelength periodicity of the dense meander inhibits propagating transmission into free space via evanescent wave confinement. This combined mechanism drives efficient Ohmic dissipation within the metallic meander, yielding an experimental absorption efficiency ranging from 50% to 90% across the 12.5-30 $\mu$m wavelength range. The cavity-free design greatly simplifies the focal plane array fabrication while providing the broadband response required for the next generation detectors. To assess the suitability of the $Ti_{40}V_{60}$ alloy as an active superconducting detector material, a test LEKID resonator was also designed, fabricated, and experimentally characterized. An RF resonance at frequency close to the simulated response was observed at temperatures below superconducting transition, thereby confirming the suitability of Ti-V alloys for LEKID development.

# 1. Introduction

The electromagnetic spectrum spanning the 10 to 50 $\mu$m wavelength range, bridging the MIR and the onset of the FIR, is a critical observational window for both astrophysics and atmospheric science. In astronomy, this regime is crucial for characterizing cold astrophysical objects, probing protoplanetary disks, and identifying potential biosignatures such as methane, ozone, and nitrous oxide in the atmospheres of terrestrial exoplanets [1,2]. On the earth, it is equally important for atmospheric spectroscopy and continuous climate monitoring [3,4]. Despite its importance in several fields, the progress across this spectral band has been constrained by the lack of fast, ultra-sensitive, and easily multiplexable detectors [5].

Conventional detector technologies in this MIR-FIR transition regime face significant physical and practical limitations. Semiconductor detectors such as HgCdTe exhibit rapidly increasing dark current and reduced sensitivity as the cutoff wavelength approaches and exceeds 15 $\mu$m [6,7]. Blocked Impurity Band (BIB) detectors, including Si:As and Si:Sb, currently remain the standard at longer wavelengths and are employed in the MIR studies. However, they are difficult to scale into large-format arrays and remain constrained by their intrinsic read noise [8-10]. Thermal detectors such as bolometers and Transition Edge Sensors (TES) can offer excellent sensitivity. However, bolometers suffer from inherently slow thermal response times [11], while the TES arrays require complex and power-dissipating cryogenic multiplexing systems such as SQUID readouts, which complicate instrument architecture and hinder megapixel scalability [12,13]. Consequently, a detector capable of combining the single-photon sensitivity, fast response, and straightforward large-array multiplexing across the 10-50 $\mu$m band remains an urgent requirement.

Superconducting Microwave Kinetic Inductance Detectors (MKIDs) have emerged as a leading technology to meet these requirements [12]. MKIDs operate on the principle of the superconducting pair-breaking phenomenon. Photons with energies greater than the superconducting energy gap generate quasiparticles, thereby modifying the complex surface impedance and kinetic inductance of a superconducting film embedded in a high-quality factor microwave resonator [14]. Because each resonator can be tuned to a unique frequency, thousands of detectors can be read out simultaneously through a single microwave feedline using frequency-division multiplexing, thus eliminating the wiring bottlenecks of earlier detector technologies [12,14]. Among the MKID variants, the LEKID is particularly attractive for high-density arrays [15]. By combining the radiation absorber and inductive meander into a single continuous structure, the LEKIDs avoid complex antenna coupling schemes, simplify fabrication, permit polarization engineering when desired, and improve the focal plane filling fractions [16,17].

LEKIDs have already reached high technological maturity in the sub-millimeter and deep FIR regimes (typically >100 $\mu$m), as demonstrated by instruments such as NIKA/NIKA2 and concepts for future space missions [18,19]. Their extension into the shorter 10-50 $\mu$m MIR-FIR band, however, remains largely unexplored [20]. At these wavelengths, efficient optical coupling and superconducting thin-film optimization become significantly more demanding. In conventional LEKID architectures, the inductive meander serves as both the microwave resonator and the direct optical absorber. High absorption efficiency is typically achieved by designing the LEKID meander so that its effective impedance, determined by the normal-state sheet resistance and the geometry of the superconducting film, matches that of free space. The absorber is then suspended above a perfectly reflecting metallic backshort, positioned at a quarter-wavelength (λ/4) distance, this reflector forms a resonant cavity that maximizes the electric field at the absorber plane [21]. While highly effective at millimeter wavelengths, this approach becomes increasingly impractical in the MIR regime. For example, a λ/4 cavity designed for a 20 $\mu$m wavelength requires a suspended gap of only 5 $\mu$m, making fabrication of uniform cavities across large detector arrays extremely challenging and yield-limiting. In addition, the use of a λ/4 backshort cavity inherently results in narrowband absorption [22], making it unsuitable for continuous ultra-broadband MIR/FIR spectroscopy and photometry. To overcome these limitations, sub-wavelength metamaterial absorbers and plasmonic structures have been widely explored in room-temperature nano-photonics [23,24] . By tailoring metallic geometries, such structures can couple the incident radiation into localized surface modes and achieve strong absorption without reflective cavities [25]. However, translating these concepts into superconducting LEKIDs remains nontrivial, since the absorbing meander must simultaneously maintain a high-quality-factor microwave resonance while remaining compatible with different substrates and device-fabrication processes. Only recently has the KID technology begun advancing into the 20-25 $\mu$m regime: The recent milestones include the demonstration of single-photon sensitive KIDs operating at 25 $\mu$m [20] and dual-polarization MIR KIDs at 20 $\mu$m for high-sensitivity astronomical imaging [26]. Nevertheless, lumped-element architectures specifically engineered for the 10-50 $\mu$m band remain exceedingly scarce.

The targeted wavelength range of such MIR/FIR LEKIDs directly overlaps with the tuning range of the Infra-Red Free Electron Laser (IR-FEL) at the Raja Ramanna Centre for Advanced Technology (RRCAT), which delivers tunable coherent radiation from 12.5 to 50 $\mu$m [27]. This creates a valuable alignment between detector development and source-based characterization. On one hand, the exceptional sensitivity and fast response of the LEKIDs make them promising detectors for IR-FEL-based user experiments involving time-domain spectroscopy and advanced materials studies under varying magnetic fields and temperatures, as these applications often require detection of weak signals generated by highly absorbing samples irradiated by picosecond FEL micro-pulses. The detection of these weak signals are beyond the practical limits of conventional semiconductor sensors. On the other hand, the IR-FEL provides an ideal high-brightness platform for systematic characterization of the newly developed LEKIDs. Its continuous wavelength tunability enables direct measurement of wavelength-dependent absorption efficiency. Furthermore, since the RRCAT IR-FEL employs a planar pure permanent magnet undulator, the emitted radiation is intrinsically linearly polarized [27]. Since the meander geometry of a standard LEKID couples strongly to electric fields aligned parallel to its traces, these structures have high polarization selectivity [28]. This property is particularly valuable for evaluating the polarization response of the detectors under test.

Motivated by these scientific and technological opportunities, this work presents the design, fabrication, and characterization of a novel cavity-free LEKID-compatible absorber architecture optimized for the 12.5-50 $\mu$m wavelength range. By engineering the lumped-element meander geometry and selecting $Ti_{40}V_{60}$ superconducting thin films, broadband response and substantially high optical absorption are achieved without using conventional reflective cavities. A test LEKID resonator is also designed and experimentally characterized to verify the superconducting resonator functionality of the $Ti_{40}V_{60}$ alloy. These results establish $Ti_{40}V_{60}$-based MIR-FIR LEKIDs as a promising, scalable, and ultra-sensitive option for next-generation instrumentation.

## 2. Experimental Methods and Electromagnetic Simulation

To examine the broadband optical absorption of $Ti_{40}V_{60}$ metallic meanders and assess the potential of the Ti-V alloy thin films as an active superconducting element for the LEKIDs, two sets of devices with different geometries were fabricated: (i) three meander absorbers with similar line widths (*LW*: 6 $\mu$m) but varying line spacings (*LS*: 3, 6, and 12 $\mu$m) intended for MIR absorption measurements, and (ii) a test LEKID resonator for microwave characterization at temperatures down to 1.5 K. The $Ti_{40}V_{60}$ thin films were deposited using a home-built DC and pulsed-DC magnetron system by co-sputtering high-purity Ti and V targets (M/s Testbourne, U.K., with purities of 99.99% and 99.9% respectively). Details of the deposition system and the growth parameters for precise control of the alloy stoichiometry are provided elsewhere [29-31]. The films were grown on Si (100) substrates (10 mm × 10 mm × 0.5 mm). The deposition was carried out at room temperature in UHP Ar (99.9995%) at a background pressure of $0.63 \times 10^{-3}$ mbar, after evacuating the deposition system to a base pressure better than $8 \times 10^{-8}$ mbar. Alloy composition was controlled through co-sputtering rates calibrated as a function of sputtering time and sputtering currents of the individual targets [29,30,32]. Prior to the deposition, the substrates were cleaned in de-ionized water, boiling acetone, and ethyl alcohol, followed by drying at 100 °C. To achieve uniform thickness and alloy composition, the substrates were rotated at 25 rpm during the deposition.

The meander patterns were lithographed on Si substrates using a home-built direct laser writer-based UV photolithography system. The system uses a tightly focused 405 nm diode laser and computer-controlled xy-raster stages to expose photoresist-coated substrates according to soft masks designed with the open-source software Klayout (Layout Editor; KLayout 0.28.7) and Phidl, [33] a Python-based device layout tool. After exposing and developing the photoresist, the $Ti_{40}V_{60}$ films were deposited, and the undesired material was removed through lift-off using acetone as the solvent. Three meander geometries were fabricated with a consistent line width of 6 $\mu$m, while the line spacings varied at 12 $\mu$m, 6 $\mu$m, and 3 $\mu$m. To achieve samples with the desired effective sheet resistance ($Z_{meander}$), the film thickness was adjusted from 20 nm for the $LS$ = 12 $\mu$m sample to 10 nm for the $LS$ = 3 $\mu$m sample. The sheet resistance ($R_S$) of the films was estimated from room-temperature resistivity measurements conducted on the control samples patterned in the Hall bar geometry, deposited alongside the meanders.

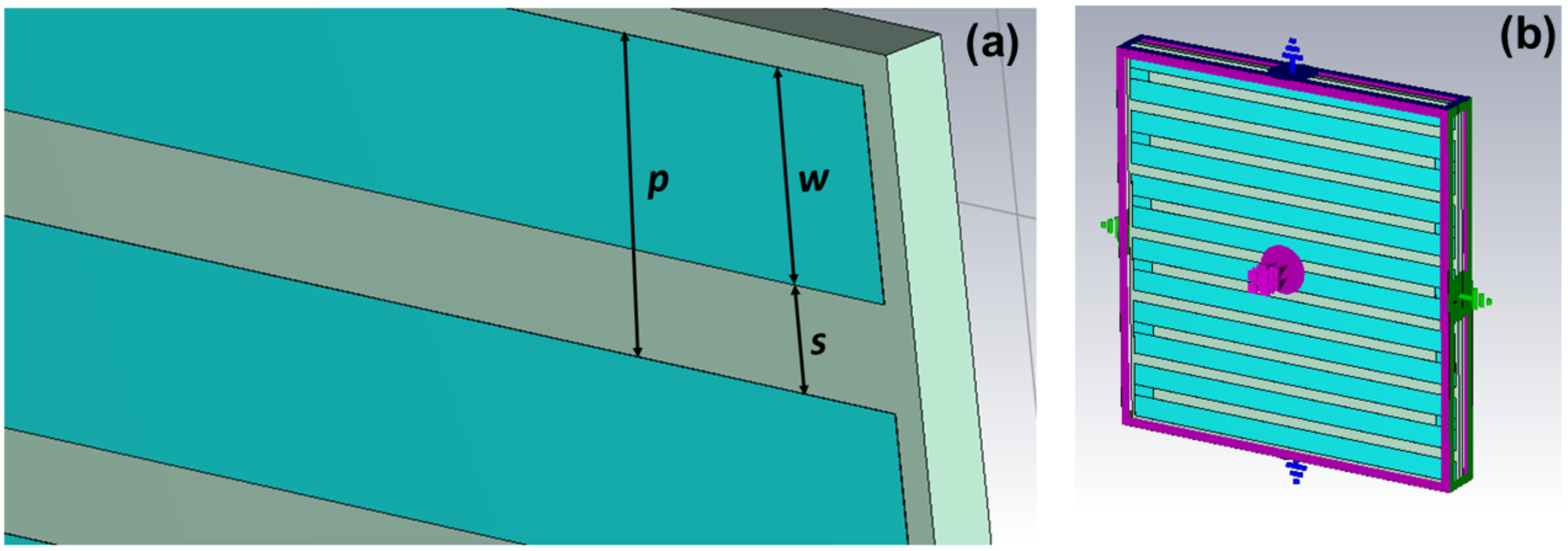

**Fig. 1 (a)** Zoomed-in image of the schematic meander structure illustrating the meander. *p*, *w*, and *s* represent the pitch, width, and separation of the meander lines. **(b)** Schematic diagram of the meander structure on the substrate used for CST simulation. The left and right boundaries (green) were assigned as perfect electric conductor (PEC) boundaries, while the upper and lower boundaries (blue) were defined as perfect magnetic conductor (PMC) boundaries. The ports normal to the absorber plane (purple) were set as open boundaries for excitation of the IR radiation along the meander lines.

Electromagnetic simulations of the absorber structures were performed using CST Microwave Studio over 12.5-50 $\mu$m spectral range. The lossy silicon was used as the substrate material, and the sheet resistance of the control Ti-V alloy samples were used to simulate the effective sheet resistance of the corresponding meanders. Fig. 1(a) shows the meander geometry, where *p*, *w*, and *s* denote pitch, line width (*LW*), and line spacing (*LS*), respectively. The CST optical simulation model, along with the boundary conditions, is shown in Fig. 1(b). In order to simulate the linearly polarised IR-radiation along the meander lines, the left and right boundaries were assigned as perfect electric conductors (PEC), while the upper and lower boundaries were defined as perfect magnetic conductors (PMC). The ports normal to the absorber plane were set as open boundaries. To reproduce the experimental conditions, the incident wave was introduced from the substrate side towards the metallic meander. The absorption in the metal layer was then calculated using the simulated $S_{11}$ and $S_{21}$ parameters.

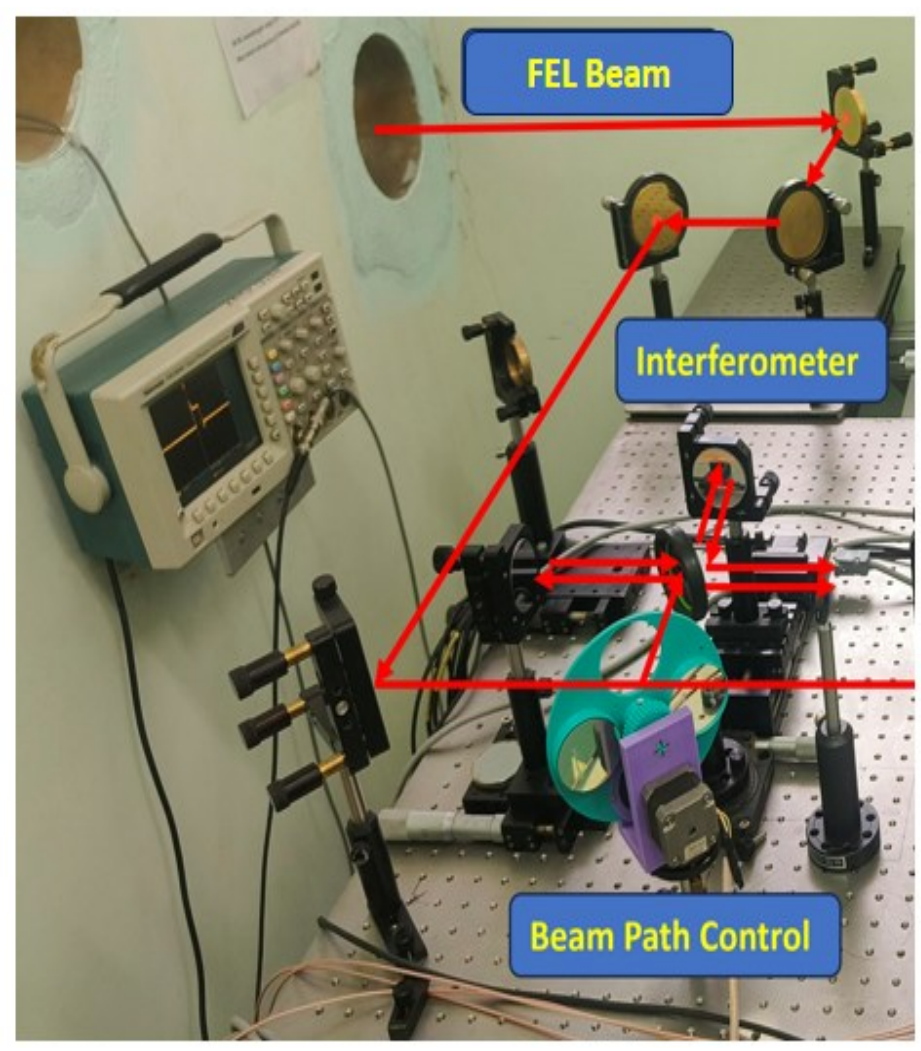

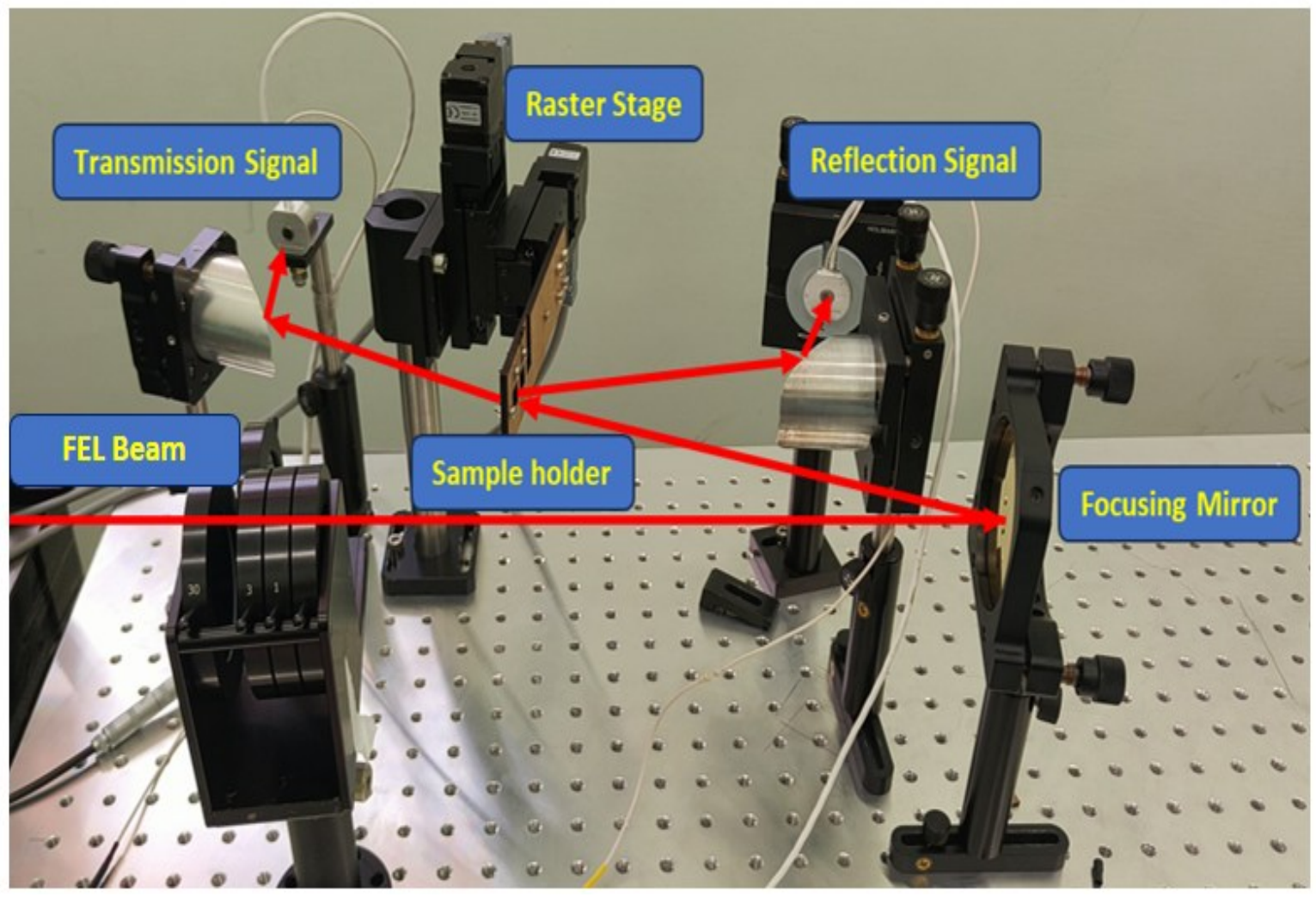

**Fig. 2** Photographs of the experimental optical measurement setup used for simultaneous transmission and reflection characterization of the fabricated meander absorber structures using the IR-FEL (right side). Red arrows represent the direction and path of the IR-FEL beam. The setup includes beam steering mirrors, focusing optics, pyroelectric detectors, and a home-built Michelson interferometer (left side) for independent wavelength determination.

Optical transmission and reflection measurements were performed on four devices across the wavelength range of 14-30 $\mu$m, using the optical setup built around port P-4 of the IR-FEL facility at RRCAT. Photographs of the experimental setup is shown in Fig. 2. Red arrows represent the direction and path of the IR-FEL beam. The IR beam was focused onto the sample plane with a 250 mm focal-length, gold-coated concave mirror (75 mm diameter). The 60 mm-diameter incident IR beam was focused to a ~200 $\mu$m point and was incident on the samples at a near-normal angle. The composite device (meander+Si substrate), an Al-coated Si wafer serving as a reference for reflectivity measurements, and an aperture to collect the reference IR intensity were mounted on a computer-controlled xy-raster stage. Reflection and transmission signals from the device and references were collected using two 90° off-axis parabolic mirrors with 50 mm focal length. Two M/S WiredSense, GmbH-made SPY pyroelectric detectors (SPY-01-0022) simultaneously recorded the transmission and reflection signals. The transmission and reflection intensities of the reference substrate were obtained from the same Si substrate adjacent to the meander's 3 mm × 3 mm (for $LS$=12 $\mu$m) and 2 mm × 2 mm (for $LS$=3 $\mu$m) area. Additionally, the reflectivity and transmittivity of the substrate were estimated using the reflected intensity from an Al-coated mirror and the transmitted intensity through the aperture, respectively. The exact wavelength of the incident IR radiation was determined using a home-built Michelson interferometer [22]. Absorption in the metallic meander was estimated after de-embedding the effect of Si substrate from the transmission and reflection data measured on the composite device. The transmission and reflection images of the samples were collected using the same setup. The focussed IR-beam was used to irradiate different parts of the samples by rasterising the sample in the XY-plane while simultaneously collecting the transmitted and reflected intensities at each 0.2 mm x 0.2 mm spot. The ImageJ software [34] was used to generate the false colour images from the 2D-intensity matrix. To reduce the pixelation in the as collected images, the pixel map was upscaled 10 times using bicubic interpolation.

To assess superconducting resonator performance, a test LEKID was fabricated from the same $Ti_{40}V_{60}$ film on a 500 $\mu$m thick Si substrate. The resonator layout was designed using Phidl, a Python-based device layout tool [33]. A 400 $\mu$m-wide microstrip feedline was used to achieve a 50 Ω impedance for proper matching with the measurement system, while the ground plane was formed from the same $Ti_{40}V_{60}$ film, resulting in a complete microstrip configuration. Both the resonator and ground plane had a thickness of 40 nm. The inductor had a line width of 15 $\mu$m and spacing of 100 $\mu$m (pitch = 115 $\mu$m).

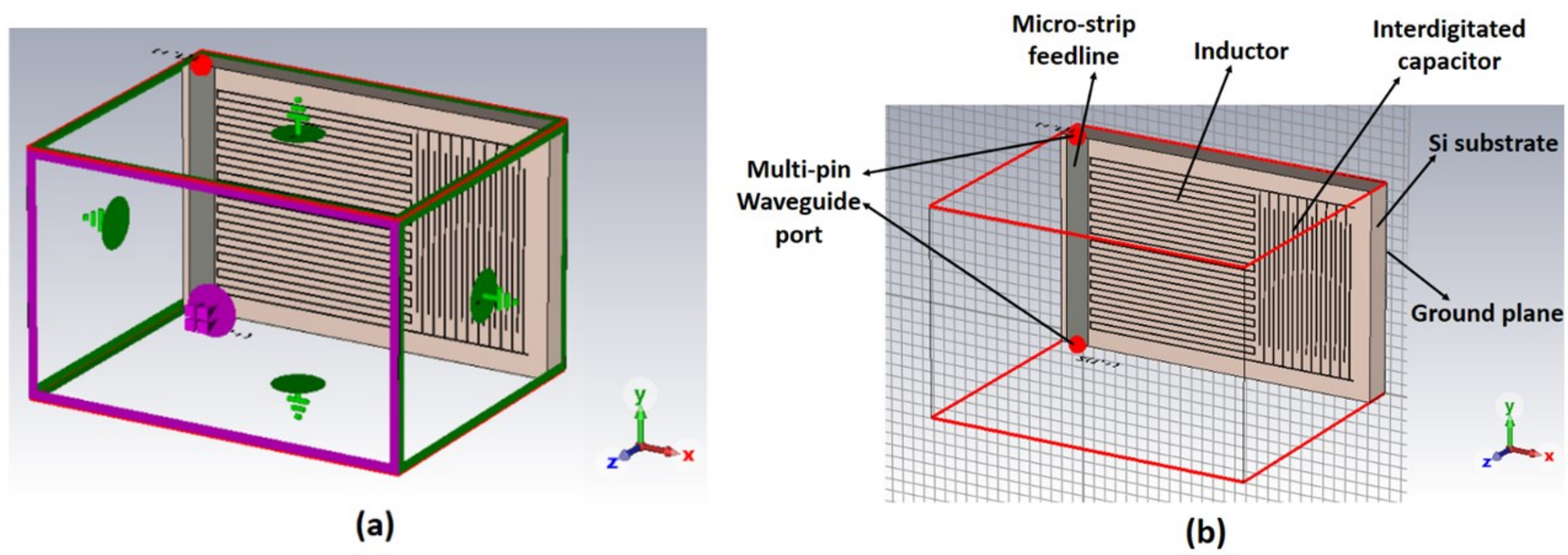

**Fig. 3 (a)** CST simulation model of the LEKID resonator showing the electromagnetic enclosure, applied boundary conditions, and excitation through the multi-pin waveguide port. **(b)** Detailed view of the LEKID geometry illustrating the microstrip feedline, meandered inductor, interdigitated capacitor, silicon substrate, and ground plane.

Prior to the fabrication, the device, including the resonator, feedline, substrate, ground plane, and the surrounding enclosure, was modelled in CST Microwave Studio, as shown in Fig. 3(a). A detailed view of the resonator geometry is presented in Fig. 3(b). The side walls and top/bottom boundaries were defined as perfect electric conductors, while the boundaries normal to the resonator axis were assigned open conditions to suppress artificial reflections.

Excitation was applied through a multi-pin waveguide port connected to the feedline. The simulated scattering parameters were used to predict the resonance frequency and optimize the coupling configuration before fabrication. The fabricated LEKID was characterized in M/s Cryogenic Inc., U. K. make cryostat using a VNA (Rohde & Schwarz, ZNL 6) by measuring the temperature-dependent microwave transmission response around the resonance. Comparison of the simulated and measured responses was used to evaluate $Ti_{40}V_{60}$ as a candidate materials-platform for LEKID applications.

# 3. Results and Discussion

Fig. 4 shows the optical microscope images of the three meander structures. The meanders were fabricated at the centre of 10 mm x 10 mm Si substrates. For these absorbers, the pitch-to-line-width (*p/w*) ratios were 3, 2, and 1.5 for line spacings of 12, 6, and 3 $\mu$m, respectively. For efficient radiation coupling into the meander absorber, the impedance of the absorber needs to match that of the medium of the incident radiation. Accordingly, the effective sheet resistance of the meander, described by the relation [16], $\left(Z_{LEKID} = R_S \times \frac{p}{w}\right)$ should be equal to the impedance of the medium in which the incident wave is traveling. The $Ti_{40}V_{60}$ films were deposited with the required intrinsic sheet resistance ($R_S$) values so that the effective sheet resistance of each patterned geometry approached the wave impedance of silicon (~110 Ω). In order to understand the effect of meander geometry on broadband absorption and validate the proposed cavity-free absorber concept, the measured optical response of each structure was compared with electromagnetic simulations. Since the fabricated devices were realized on Si substrates, the measured transmission and reflection responses include not only the contribution of the metallic meander layer, but also optical effects originating from the thick silicon substrate. Therefore, prior to comparing the intrinsic absorber performance of the three structures, an optical de-embedding procedure was employed to isolate the true absorption of the $Ti_{40}V_{60}$ patterned metal layer.

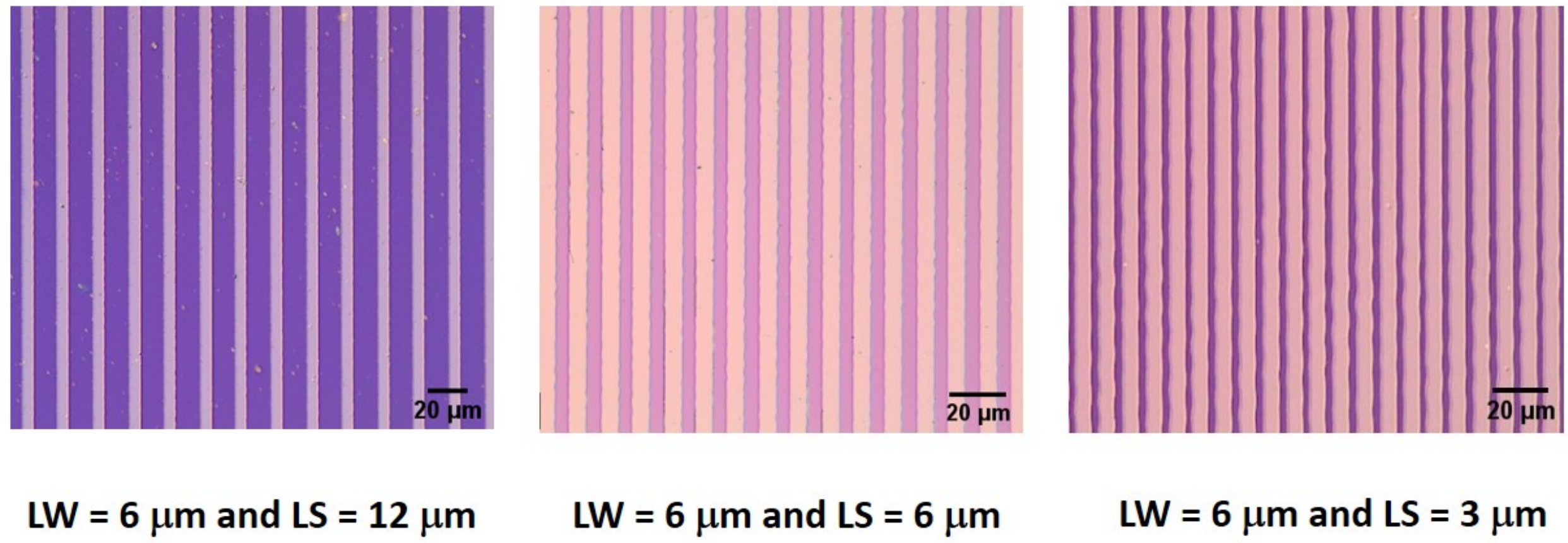

**Fig. 4** Optical microscope images of patterned structures with a constant line width (*LW*) of 6 $\mu$m and varying line spacings (*LS*). The lengthscale for each image is indicated within the panels.

To accurately determine the intrinsic optical absorption of the $Ti_{40}V_{60}$ patterned layer, the contributions of the thick silicon substrate, namely front-surface reflection and bulk internal absorption, were mathematically de-embedded from the measured device response. Since the optical round-trip distance inside the substrate (*2nd = 3.42 mm; n* = Si refractive index and *d*= substrate thickness in mm) exceeds both the optical wavelength and the source coherence length ~ 3 mm (for 10 pico-second micro-pulse width of IR-FEL light) [27] the interference between incident beam and the multiple reflected beams within silicon does not occur. Additionally, to avoid the spectral modulation due to interference effects arising from multiple reflections, the IR beam was incident at a near normal angle (less than $10^0$) to the substrate. Accordingly, the light propagation through the substrate was modelled using an incoherent multiple-reflection framework based on Airy intensity summation [35]. Conversely, the nanometre thick $Ti_{40}V_{60}$ film was treated as a single effective optical boundary whose macroscopic interface properties inherently capture the coherent wave interactions within the thin metallic layer [36]. The optical de-embedding procedure first requires calibrating the intrinsic properties of the bare silicon wafer. Using the experimentally measured near-normal-incidence reflectance ($R_{sub}$) and transmittance ($T_{sub}$) of the substrate, we isolated the single-surface reflectivity of the Air-Silicon interface ($R_{Si}$) and the single-pass internal bulk transmittance ($\tau_{bulk}$). By algebraically inverting the geometric series of internal reflections for a symmetric, thick dielectric slab, the $R_{Si}$ is found as the physically valid root ($0 < R_{Si} < 1$) of the equation (1) [37,38].

$$(R_{sub} - 2)R_{Si}^{2} + \left(T_{sub}^{2} - R_{sub}^{2} + 2R_{sub} + 1\right)R_{Si} - R_{sub} = 0 \qquad (1)$$

The corresponding internal bulk transmittance, which account for the intrinsic absorption of the Si medium per pass, is then extracted from the intensity conservation identity [35]

$$\tau_{bulk} = \frac{R_{sub} - R_{Si}}{R_{Si} T_{sub}} \tag{2}$$

For the full device structure (air/silicon/metal structure) where the sample is illuminated from the substrate side [Fig. 2(b)], the total measured reflectance ($R_{dev}$) and transmittance ($T_{dev}$) include contributions from the front air/Si interface, internal propagation $\tau_{bulk}$, and the back Si/metal interface. We define $R_{int}$ and $T_{int}$ as the phenomenological reflectance and transmittance of the Silicon-Metal boundary, respectively, as evaluated from within the silicon medium. By expanding the incoherent sum of intensities for this asymmetric cavity and rearranging for the target interface terms, we decouple the metal's response from the substrate [39,40] using,

$$R_{int} = \frac{R_{dev} - R_{Si}}{\tau_{bulk}^{2}[(1 - R_{Si})^{2} + R_{Si}(R_{dev} - R_{Si})]} \tag{3}$$

$$T_{int} = \frac{T_{dev}[1 - R_{Si} R_{int} \tau_{bulk}^{2}]}{(1 - R_{Si}) \tau_{bulk}} \tag{4}$$

With the substrate effects rigorously removed, the macroscopic optical properties of the metal film are isolated. The intrinsic absorption fraction of the $Ti_{40}V_{60}$ layer was then found as the residual of the reflected and transmitted intensities of the decoupled interface:

$$A_{metal} = 1 - R_{int} - T_{int} \tag{5}$$

This treatment eliminates the artificial enhancement caused by multiple secondary passes through the absorbing silicon substrate, ensuring that the extracted spectra represent the true optical dissipation within the $Ti_{40}V_{60}$ meander layer.

Figure 5 (a)-(c) shows the measured and simulated absorption spectra of the three $Ti_{40}V_{60}$ meander absorbers having a fixed line width of 6 $\mu$m and line spacings of 12, 6, and 3 $\mu$m respectively. All the simulation parameters were kept consistent with the experimental parameters. The simulations were performed for the full wavelength range (12.5 - 50 $\mu$m) of the IR-FEL. A clear geometry dependent enhancement in the broadband absorption is observed, with progressively denser meanders delivering noticeably stronger optical dissipation. Moreover, the features in the simulated and experimental absorption spectra broadly matches. Since the meanders were much larger as compared to the ~200 $\mu$m focussed IR-beam size, the absorbers can be treated as metal gratings. As per the Drude theory [41-43], the plasmon resonance frequency of the metal gratings is given by,

$$\lambda_P = \frac{2\pi c}{\omega_P} = \sqrt{\pi^2 ps} \qquad (6)$$

Where, $p$ is the period and $s$ is the separation between the grating lines. The transmission spectra show peak at the plasmon resonance frequency, which as per the equation (6), red shifts with increase in the period or spacing of the grating. The peaks in the transmission spectra corresponds to the valleys in the absorption spectra. For our absorber structures, the absorption valleys shall occur at 46.14 $\mu$m, 26.64 $\mu$m and 16.31 $\mu$m for $LS$ = 12, 6 and 3 $\mu$m respectively. As evident from the figure 5 (d), the minima in the simulated absorption spectra of all the three absorbers occur nearly at the calculated plasmon resonance wavelengths. The $\lambda_P$ for the $LS$ 3 and 6 $\mu$m samples fall within the experimental wavelength range and the absorption minima in the measured absorption spectra matches closely with that of the simulated one.

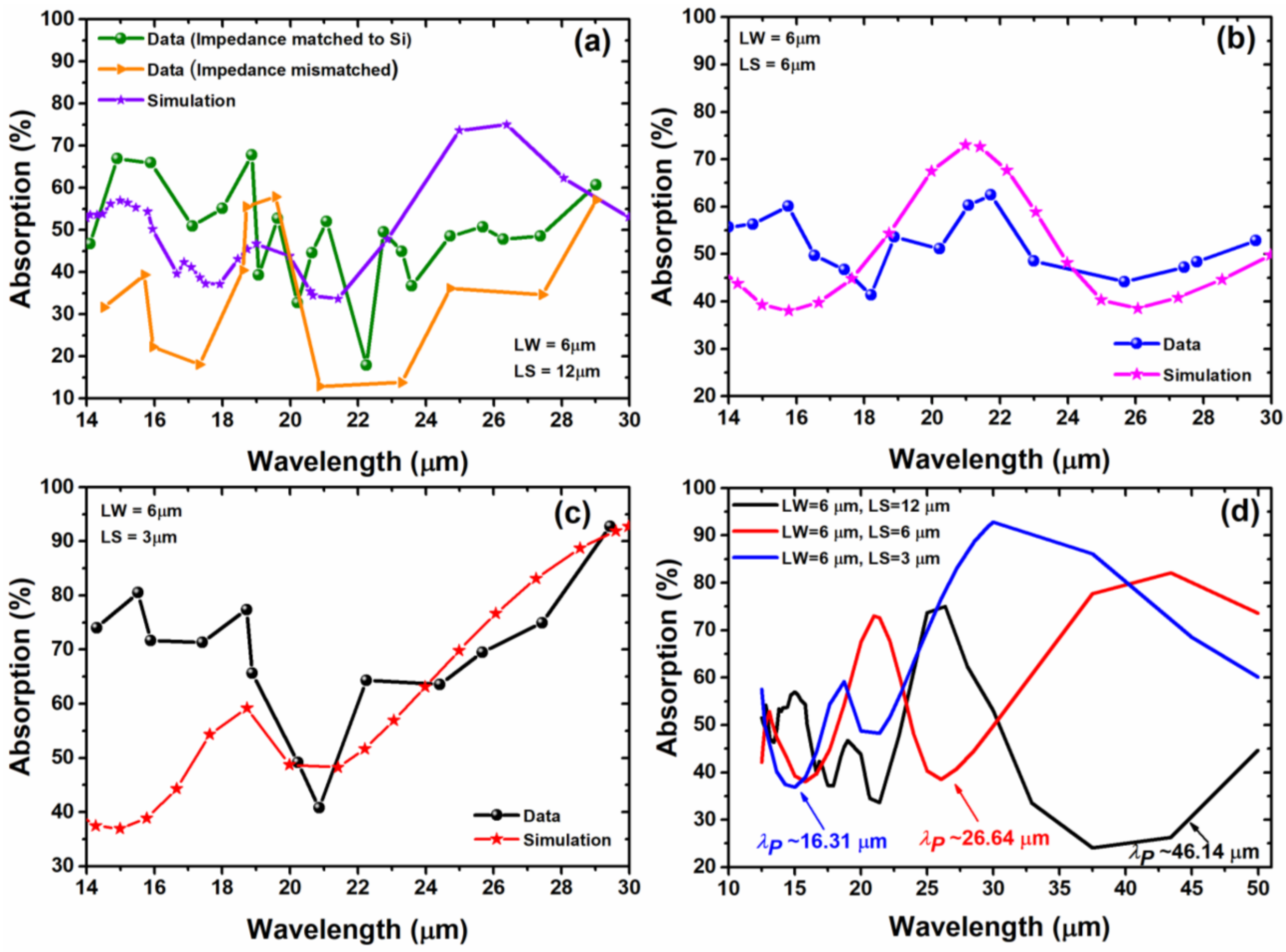

**Fig. 5** Comparison between experimentally measured (Data) and simulated (Simulation) absorption spectra as a function of wavelength for meander absorber structures with a fixed line width ($LW$ = 6 $\mu$m) and different line spacings ($LS$). (a) Absorption spectra for $LS$ = 12

$\mu$m, showing measured absorption for the impedance-matched and impedance-mismatched meander structures, figure also shows the simulated absorption curve. (b) Comparison of measured and simulated absorption spectra for $LS$ = 6 $\mu$m. (c) Comparison of measured and simulated absorption spectra for $LS$ = 3 $\mu$m. (d) Simulated absorption spectra of all three structures ($LS$ = 12, 6, and 3 $\mu$m) over the extended wavelength range of 12.5-50 $\mu$m, showing the shift in the $\lambda_P$ with decreasing line spacing.

For the structure having $LS$ = 12 $\mu$m, the measured response is broadband but relatively non-uniform, with absorption typically in the range of ~35-70%. A broad maximum appears near 18-20 $\mu$m, where absorption approaches ~70%, followed by a dip around 22 $\mu$m. At longer wavelengths, the absorption recovers to ~50-60%. Simulations reproduce the overall broadband behaviour, although local deviations remain, which, we believe, might be due to the power fluctuations inherent to the radiation generated in the IR-FELs. To confirm the experimental observations as well as impedance matching criterion, a fourth sample having $LW$ 6 $\mu$m and $LS$ 12 $\mu$m was fabricated with slightly lower sheet resistance. The absorption spectrum of this sample is also shown along with the absorption spectrum of the sample having matched impedance in figure 5 (a). The overall features of the absorption spectrum remain the same for both samples. However, the absorption efficiency of the unmatched absorber remains lower than that of the matched meander having $Z_{meander}$ ~ 109 Ω. Since, this geometry has the lowest metallic filling fraction, coupling between the incident field and resistive current paths is comparatively weaker.

Reducing the spacing to 6 $\mu$m produces a clear improvement in both absorption magnitude and spectral uniformity. The pronounced minima observed in the previous design are strongly suppressed, and absorption remains within ~40-70% over most of the measured wavelength band. At longer wavelengths, the absorption further increases to ~75%. The simulated spectrum follows the same trend, confirming that increased metal coverage and the reduced inter-line gaps enhance the confinement of the incident electromagnetic field within the meander structure. As a result, a larger fraction of the optical energy interacts with the $Ti_{40}V_{60}$ film, leading to increased resistive (Ohmic) dissipation and consequently higher absorption. The best performance is obtained for the densest structure with 3 $\mu$m line spacing. In this case, absorption remains above 50% across the entire measured wavelength range and rises sharply at longer wavelengths, finally exceeding 90% near the long-wavelength edge. The simulated spectrum closely follows the measured trend and likewise predicts near unity absorption. This is the central result of the present work, a simple $Ti_{40}V_{60}$ planar meander can realize extremely high broadband MIR absorption without employing λ/4 backshort cavities, suspended membranes, or backside reflectors.

In addition to the spectral measurements performed at single 200 $\mu$m spots within the meander structures, the optical absorption of the fabricated structures was also verified through imaging experiments. The transmission and reflection images were collected simultaneously. Representative images obtained at 26 $\mu$m wavelength for the least and most dense meander geometries, corresponding to line spacings of 12 $\mu$m and 3 $\mu$m, are shown in Fig. 6(a) and (b) respectively. Since the meander features were very small as compared to the imaging resolution of 200 $\mu$m, the meander can be seen as a 3 mm × 3mm square (for $LS$ = 12 $\mu$m) and 2 mm × 2mm square (for $LS$ = 3 $\mu$m) at the centre of the substrate. In both the absorbers, the patterned meander region appears as a dark square in the transmission images, indicating strong attenuation of the incident radiation within the absorber-covered area. The reduction in the intensity of the transmitted beam from the sample alone is not the indication of the absorption in the sample. A corresponding reflection image is required to confirm the same. If the reflection image shows darker meander as compared to the remaining substrate, then the reduction in intensity is due to absorption in the meander. Otherwise, the meander should be brighter as compared to the adjoining substrate. In our experiments, a clear contrast is observed in the corresponding reflection images confirming that the patterned metallic layer significantly modifies the local optical impedance and leads to a lower reflected intensity. Additional imaging measurements were performed for the $LS$ = 3 $\mu$m geometry at 19 $\mu$m and 29.5 $\mu$m wavelengths. These wavelengths correspond to the comparatively lower-absorption and near-maximum absorption regions of the measured spectrum respectively. At 19 $\mu$m, the patterned region exhibits relatively weaker contrast in reflection channel, consistent with the reduced absorption observed spectroscopically. In contrast, at 29.5 $\mu$m, the patterned meander appears significantly darker and more distinct, accompanied by enhanced reflection contrast, confirming the substantially stronger optical absorption at longer wavelengths. The clear wavelength-dependent evolution of the spatial contrast directly follows the measured absorption behaviour and provides additional visual confirmation of the broadband optical functionality of the optimized meander absorbers. Overall, the simultaneous contrast observed in both transmission ($T$) and reflection ($R$) channels demonstrates that the meander structures effectively redistribute the incident optical power through strong absorption, reduced transmission, and reflection, supporting the suitability of the proposed cavity-free geometry for infrared sensing and imaging applications. Figure 6(c) shows the transmission and reflection images of the same meander when the IR radiation is incident from the sample side. It can be observed that the reflected intensity from the meander is much higher than the adjoining Si substrate, attributed mainly to the impedance mismatch between air (377 Ω) and the meander (109 Ω). So, instead of getting absorbed, the IR radiation is simply reflected back.

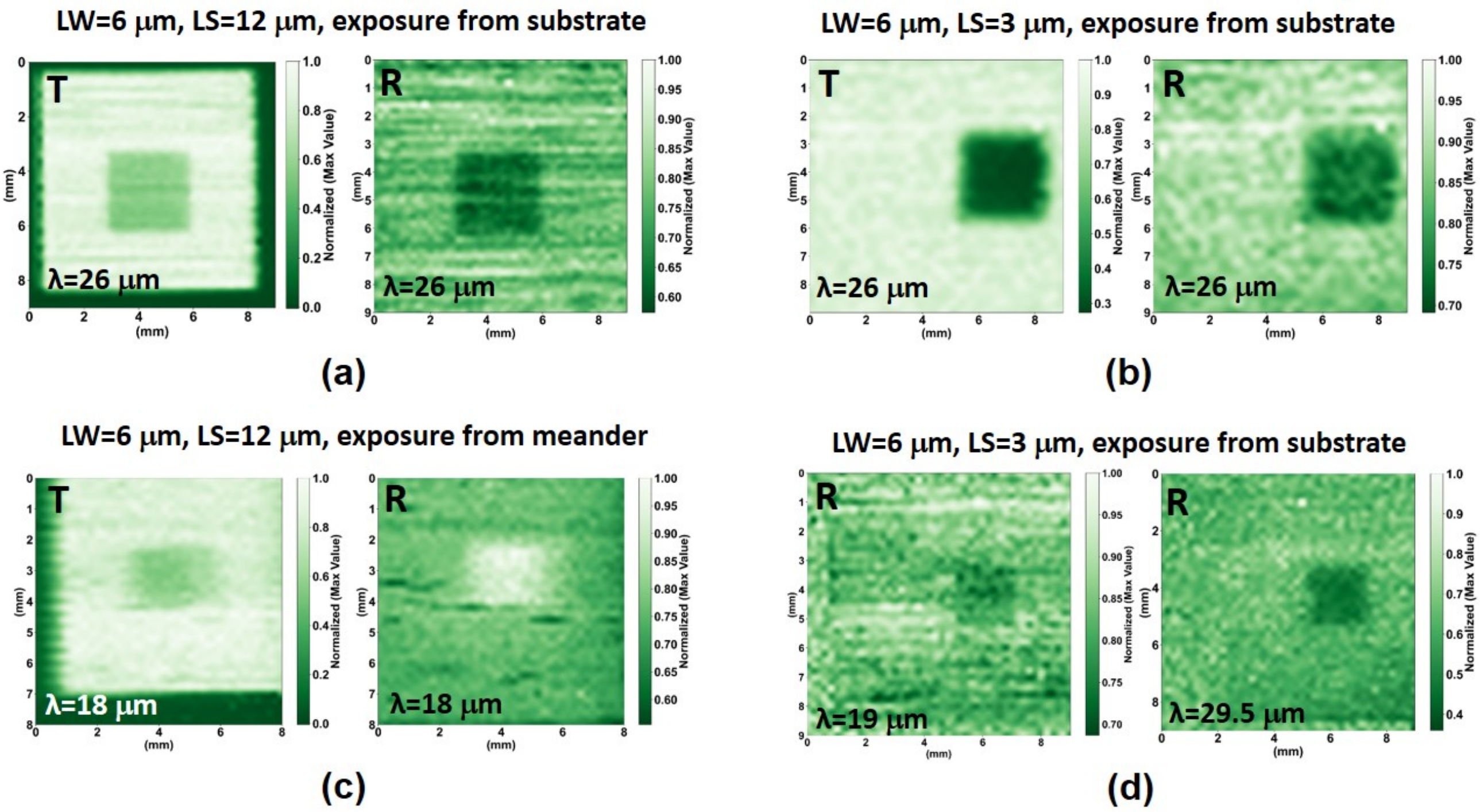

**Fig. 6** Two-dimensional spatial imaging of the fabricated meander absorbers obtained using the IR-FEL light in the transmission (*T*) and reflection (*R*) modes. **(a)** and **(b)** show the imaging results measured at a wavelength of λ = 26 μm for the structures with LW = 6 μm, LS = 12 μm and LW = 6 μm, LS = 3 μm, respectively. **(c)** presents the imaging obtained under front-side (meander) exposure, where the reflection image shows strong reflected intensity from the patterned region due to poor impedance matching in this configuration. In contrast, the other measurements were performed under back-side (substrate) illumination through the Si substrate, providing improved impedance matching and enhanced absorption contrast. **(d)** shows the reflection images of the LW = 6 μm, LS = 3 μm structure measured at λ = 19 μm and λ = 29.5 μm, corresponding to the comparatively lower-absorption and higher-absorption regions of the measured spectrum, respectively. All the images are normalized between 0 and 1, where brighter regions correspond to higher transmitted/reflected intensity and darker regions indicate reduced intensity due to stronger absorption by the patterned meander absorber.

The exceptionally high absorption (up to 90%) observed in the metallic meanders, without any metallic back-short, deviates significantly from fundamental theoretical limits expected for conventional thin resistive films. For a freestanding metallic sheet without back short (in a symmetric environment: air/metal/air), the theoretical maximum absorption limit is 50% [21]. The experimentally observed ~90% absorption therefore cannot arise from ordinary sheet losses alone. It indicates to an additional confinement-assisted absorption pathway enabled by the synergistic coupling mechanism involving asymmetric impedance matching, evanescent-wave trapping and possible excitation of hybrid surface polaritons as well [44,45].

The primary requirement for efficient coupling of incident radiation is established by the asymmetric dielectric environment (air/Si/metal/air) of the meanders. Since the illumination occurs through the Si, and the $Z_{meander}$ is matched to the wave impedance of silicon (~110 Ω) [21], the zeroth order back reflections at the Si/metal interface are minimized. But, even for an impedance-matched sheet placed between silicon and air, the analytical maximum absorption limit is given by equation (7) [46],

$$A_{max} = \frac{4{n_1}^2}{(2n_1+n_2)^2} \tag{7}$$

With $n_1 = 3.42$ (Si) and $n_2 = 1$ (air), the maximum absorption in an unstructured metallic sheet in asymmetric dielectric stack is limited to only ~76%, with remaining 24% of the power getting transmitted into the air as zero-order wave.

However, unlike a continuous film, the structured meander acts as a resonant metamaterial, actively harvesting the theoretical transmitted power and also by suppressing the transmission into the free space (air). Once the electromagnetic energy reaches the meander, its transmission into the surrounding free-space air is strictly restricted by grating diffraction limits [40]. For the highest performing structure, the period remains subwavelength across the operating band. According to the grating relation (8) [47], a transmitted diffraction order can propagate into air only when $|Sin\ \theta_m| \leq 1$, where, for the grating period $p$,

$$Sin\ \theta_m = \frac{m\lambda_0}{p \times n_{air}} \tag{8}$$

Since the free-space wavelength $\lambda_0$ is larger than the meander period for the measured spectral range, the first and higher diffraction orders yield $|Sin\ \theta_m| > 1$. Consequently, no real propagation angle exists for these modes, and thus they become exponentially decaying evanescent rather than radiative ones. The metal/air interface acts as a virtual back short as the optical energy remains confined to the substrate surface, and thereby gets absorbed in the impedance-matched meander.

Having established the broadband optical absorption performance of the proposed meander architecture, it is also important to assess whether the same $Ti_{40}V_{60}$ thin film platform is suitable for superconducting detector operation. For this purpose, a test LEKID resonator was designed through simulations, fabricated, and experimentally characterized to evaluate its microwave resonance response and to extract the quality factors. As discussed in the Experimental section, the inductive meander of the present LEKID was designed with a line width of 15 $\mu$m and a line spacing of 100 $\mu$m. This geometry was intentionally chosen to keep the resonance frequency below 1 GHz, so as to keep the RF electronics simpler. The resonance characteristics were first analysed using CST Microwave Studio. The detailed simulation model, boundary conditions, and excitation geometry are shown in Fig. 3(a), where the enclosure and ports were defined to closely reproduce the experimental configuration. The simulated transmission response, presented in Fig. 7(a), shows a clear resonance dip at $f_0$ = 516.4 MHz. The inset of Fig. 7(a) shows the simulated surface current density distribution at the resonance frequency, demonstrating strong current localization within the meandered inductive section of the LEKID. The highest current concentration is observed along the meander lines, confirming that the inductive section predominantly stores the electromagnetic energy and governs the resonant behaviour of the device. This also verifies the efficient coupling between the feedline and the resonator at the designed resonance frequency. Using equation (10), the loaded quality factor was determined as $Q_L = 9.7 \times 10^3$, where $\Delta f$ is the half power bandwidth of the resonance. From equation (11), the coupling factor was obtained as $k$ = 10.3, indicating an over-coupled design. Using equation (12), the corresponding intrinsic quality factor comes out to be $Q_0 = 1.1 \times 10^5$, which reflects a low internal loss under ideal simulated conditions. Using equation (13), the coupling quality factor was calculated to be $Q_C = 1.06 \times 10^4$, indicating strong coupling between the resonator and the feedline.

$$Q_L = \frac{f_0}{\Delta f} \tag{10}$$

$$k = \frac{1-|S_{21,min}|}{|S_{21,min}|} \tag{11}$$

$$Q_0 = Q_L(1 + k) \tag{12}$$

$$Q_C = \frac{Q_0}{k} \tag{13}$$

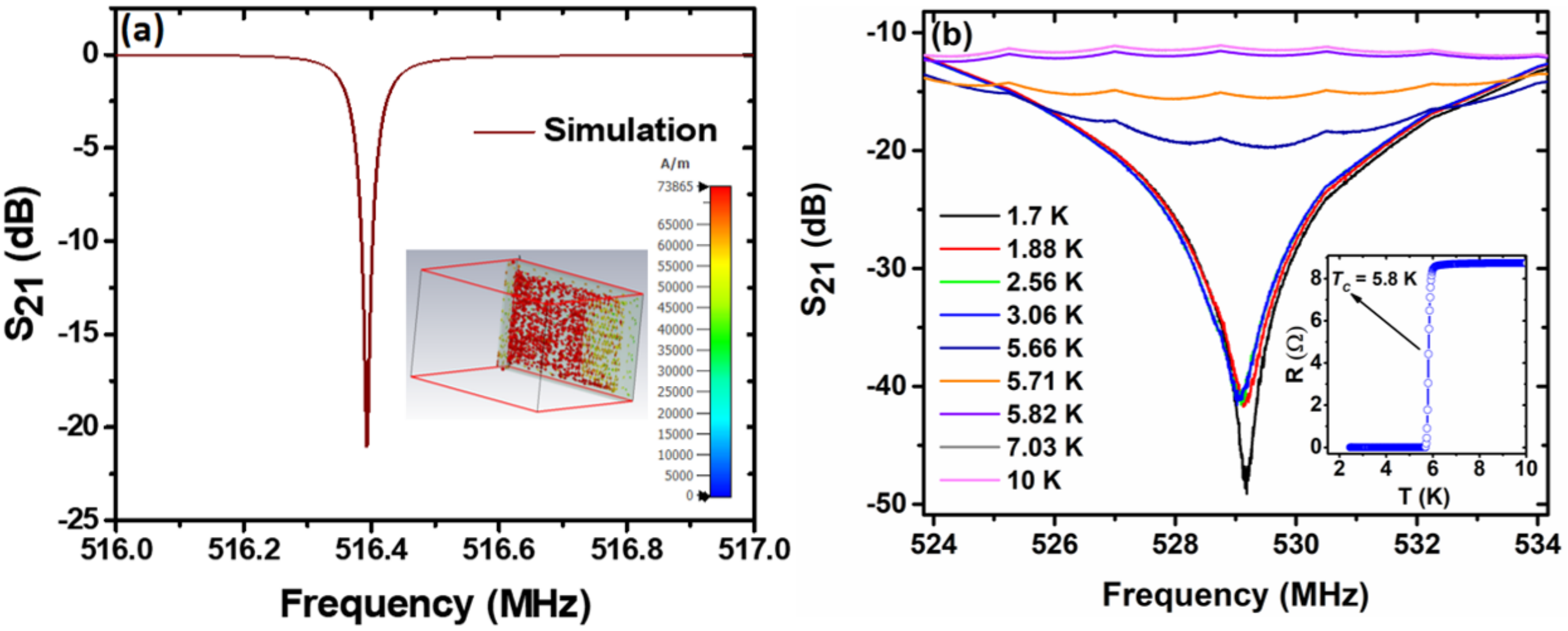

**Fig. 7 (a)** Simulated $S_{21}$ of the LEKID obtained using CST Microwave Studio, showing a resonance dip near 516.4 MHz. The inset displays the simulated surface current density distribution at the resonance frequency, demonstrating strong current localization within the meandered inductive section of the resonator. **(b)** Measured transmission response ($S_{21}$) of the fabricated $Ti_{40}V_{60}$ LEKID at different temperatures. A pronounced resonance dip is observed near 529 MHz at low temperature, which progressively broadens, weakens, and shifts toward lower frequency as the temperature approaches the $T_C$. The inset shows the resistance versus temperature, *R(T)* measurement of the same $Ti_{40}V_{60}$ thin film, indicating the $T_C$ at 5.8 K.

For microwave characterization, the device was mounted on a 50 Ω impedance-matched PCB in a 1.5 K cryostat fitted with 50 Ω coaxial cables. The lowest accessible temperature in the present setup is 1.7 K, and resonance measurements were performed from 1.7 K upward up to 10 K. Prior to the cooldown, the RF system was calibrated using standard procedures up to the PCB reference plane to remove systematic cable and connector contributions. The measured transmission response at different temperatures is presented in Fig. 7(b), while the inset shows the corresponding resistance versus temperature, *R(T)* behaviour of the same $Ti_{40}V_{60}$ thin film. At 1.7 K, a clear resonance dip is observed at $f_0$ = 529.2 MHz, in good agreement with the simulated value. As the temperature increases toward $T_C$, the resonance progressively weakens, broadens, and shifts to lower frequency, eventually disappearing near the superconducting transition, consistent with the loss of superconductivity observed in the *R(T)* curve. This behaviour is consistent with an increasing quasiparticle population, which enhances dissipation and increases the kinetic inductance, thereby degrading the resonance [21]. From the measured response, the loaded quality factor was found to be $Q_L$ = 28. The coupling factor was extracted as $k$ = 182.4, indicating a strongly over-coupled device.

The intrinsic quality factor was estimated as $Q_0 = 5.2 \times 10^3$, while the corresponding coupling quality factor $Q_C$ was obtained to be 28.3. Since $Q_C \ll Q_0$, the total loss is dominated by the coupling rather than intrinsic dissipative processes, resulting in a reduced quality factor and a broadened resonance profile compared to the simulation.

It is important to note that the present measurements are performed at a base temperature of 1.7 K, which, although well below $T_C$, is still significantly higher than the typical operating regime of LEKIDs (i.e., sub-Kelvin temperatures or $T < \frac{T_C}{10}$) [48]. As a result, a non-negligible quasiparticle population may persist in the superconducting film, leading to enhanced dissipation and a reduced intrinsic quality factor. This limitation is reflected in the experimentally obtained quality factors, where the intrinsic quality factor is considerably lower than that predicted from simulations. In the simulated case, a high intrinsic quality factor is obtained under idealized conditions with minimal losses, whereas the experimental value reflects the combined influence of residual quasiparticles, material imperfections, and practical coupling conditions. Consequently, the loaded quality factor is also significantly reduced, and the resonance appears broader as compared to the simulated response.

## Summary and Conclusion

In summary, we have demonstrated a cavity-free LEKID design architecture for broadband absorption, purely based on asymmetrically matched $Ti_{40}V_{60}$ metallic meander structures acting as metamaterials in the mid-far infrared window. By systematically varying the meander spacing at fixed line width, both simulation and experiment show that the optical response can be strongly controlled through geometrical tuning of the effective sheet impedance and metallic filling factor. Among the investigated designs, the densest meander geometry shows an experimentally measured absorption approaching 90%, achieved without metallic back reflectors, suspended membranes, or conventional quarter-wavelength resonant cavities. The high absorption originates from a combined physical mechanism where, (i) the meander sheet resistance is engineered to match the wave impedance of the silicon substrate, thereby suppressing the front-side reflection, and (ii) the subwavelength periodicity prevents higher diffraction orders from propagating into air, forcing the transmitted field into evanescent confinement near the surface. As a result, the incident optical power is efficiently converted into Ohmic losses within the metallic meander itself. An important outcome of this work is that the proposed mechanism overcomes the key limitations of the traditional LEKID optical coupling schemes. Conventional λ/4 backshort absorbers are inherently narrowband and become increasingly difficult to realize in the MIR-FIR regime because they require micron-scale cavity gaps and mechanically suspended ultrathin structures. In contrast, the present approach provides broadband high absorption over the full measured wavelength range while greatly simplifying the fabrication and large-array integration procedure.

To verify the suitability of the $Ti_{40}V_{60}$ platform for superconducting detector applications, a test LEKID resonator was also designed, simulated, fabricated, and experimentally characterized. A microwave resonance at 529 MHz was observed, which is found to be in good agreement with CST simulations. Temperature-dependent measurements confirmed the superconductivity, with a critical temperature near 5.8 K. These results demonstrate that the properly engineered $Ti_{40}V_{60}$ alloy based thin-films can simultaneously provide strong MIR absorption and functional superconducting resonator performance. Overall, this study establishes a practical and scalable pathway towards next-generation MIR-FIR superconducting detectors that combine broadband optical efficiency, simplified fabrication, and compatibility with large-arrey LEKID technology. The proposed architecture is highly promising for future focal-plane arrays, ultrasensitive spectroscopy, astronomical instrumentation, and compact infrared imaging systems.

## Acknowledgements

The authors thank Rahul Gaur (ABPS, RRCAT) and Kunver Adarsh Pratap Singh (RFSD, RRCAT) for valuable discussions and assistance with the CST simulations. Thanks are also due to Bhawani Shankar and Praveen Mohania (RFSD, RRCAT) for their help with the VNA-based RF measurements. Additionally, the authors acknowledge L. S. Sharath Chandra (FUS, RRCAT) for his help with the low temperature facility, as well as Pravin Nerpagar, Ravi Kumar Pandit, and Sarbeswar Sardar (FUS, RRCAT) for their support in the operation and maintenance of the IR-FEL machine.